\documentclass[12pt,a4paper]{article}
\usepackage[a4paper,top=2.75cm,bottom=2.75cm,left=2.85cm,right=2.85cm]{geometry}
\usepackage[T1]{fontenc}
\usepackage{lmodern}
\usepackage{amsmath,amssymb,amsthm,mathtools,bm}
\usepackage{microtype}
\usepackage{cite}
\usepackage[colorlinks=true,linkcolor=blue,citecolor=blue,urlcolor=blue]{hyperref}
\usepackage{enumitem}
\newcommand{\ap}{\alpha'}
\newcommand{\Tr}{\operatorname{Tr}}
\newcommand{\STr}{\operatorname{STr}}
\newcommand{\cD}{\mathcal D}
\newcommand{\cT}{\mathcal T}
\newcommand{\cO}{\mathcal O}
\newcommand{\cF}{\mathcal F}
\newcommand{\cH}{\mathcal H}
\newcommand{\Pinf}{P_{\infty}}
\newcommand{\Tinf}{\mathcal T_{\infty}}

\def\Tr           {\mbox{\rm Tr}\,}
\def\STr          {\mbox{\rm STr}\,}

\def\fsH    {H\!\!\!\!/\,}

\begin{document}
\begin{center}
      \Large{{Classification of Chern--Simons Couplings and Deformed Pull-Backs on D-Branes}}
\end{center}
\begin{center}
\noindent{{ Armin Hatefi}$^{a}$, 
            { Ehsan Hatefi$^{b,c,}$\footnote{ \texttt{ ahatefi@mun.ca, ehsan.hatefi@sns.it,}}}}
\end{center}
\begin{center}
\noindent{\footnotesize{$^{a}$ Department of Mathematics and Statistics, Memorial University of Newfoundland, \\
St. John’s, NL, Canada.}} \\
\noindent{\footnotesize{$^{b}$ Department of Mathematics, University of Pisa, Largo Bruno Pontecorvo 5, 56127 Pisa. }} \\
\noindent{\footnotesize{$^{c}$ Scuola Normale Superiore and INFN, Piazza dei Cavalieri 7, 56126 Pisa, Italy.}} \\
\end{center}

\begin{abstract}

We establish a geometric framework for classifying higher-derivative Ramond–Ramond (RR) couplings on Dp-branes, focusing on the complete scattering amplitude of a closed-string RR state, a gauge field, and two transverse scalars. The low-energy expansion reveals infinite massless poles and local interaction terms that are beyond standard Myers and Taylor expansions. We propose a higher-class deformation of the pullback using mixed Chern-Simons transgression operators to address these complexities. We also propose a generalized class Chern character and three conjectures: (i) the finite dimensionality of independent couplings at fixed derivative order, (ii) the representation of all-order mixed RR–gauge–scalar interactions through finitely many effective even forms, and (iii) the generation of higher-derivative contact terms from finitely many primitive tensor structures. Our framework provides a systematic mathematical foundation for reconstructing D-brane effective actions and exploring their underlying geometric and cohomological structures.
We also introduce a higher-class pullback operator \( P_{\infty} \) and mixed transgression operators \( \mathcal{T}_{\infty} \), emphasizing amplitude-derived pole residues and local contact terms as key physical data. This formalism distinguishes admissible tensor structures, presenting characteristic-form aspects as conjectures alongside their validity and their constraints.


\end{abstract}

\newpage

\section{Introduction}
\label{sec:introduction}

D$p$-branes are fundamental nonperturbative objects in superstring theory, essential for understanding its mathematical structure, duality symmetries, and phenomenological implications. Depending on the value of \( p \), corresponding to the dimensions of the D$p$-brane world volume, they appear in both type IIA and type IIB superstring theories, carrying charges under the Ramond–Ramond (RR) gauge potentials \cite{Polchinski:1995mt,Witten:1995im,Polchinski:1996na}. As dynamical sources of RR fields, D$p$-branes provide a crucial framework for studying nonperturbative aspects of string theory and the interactions between open- and closed-string sectors \cite{Hatefi:2012zh}.

The study of D-brane couplings to closed-string background fields has significantly advanced string theory, influencing the AdS/CFT correspondence, the geometric realization of gauge theories, and the microscopic description of black holes. The analysis of RR couplings \cite{Li:1995pq,Green:1996dd} has revealed connections between D-brane dynamics and nonperturbative phenomena \cite{Douglas:1995bn}. These topics have established the importance of K-theory in classifying D-brane charges \cite{Minasian:1997mm,Witten:1998cd} and led to the discovery of the dielectric effect, where the collections of D-branes become polarized in nontrivial background fields \cite{RMyers:1999ps12,Taylor:1999pr}.

Superconnection methods have also been used to develop the geometric description of D-brane charges and RR interactions. In \cite{Kennedy:1999nn} the authors investigated RR couplings on brane–antibrane systems, while  \cite{Szabo:2001yd} further connected superconnections, anomalies, and non-BPS D-brane charges as well.  The relevant low-energy theories are expressed through the Dirac–Born–Infeld (DBI) and Chern–Simons (CS) actions, which relate closely to the duality properties of superstring theory \cite{pol 12}. For discussions on these actions and their applications, see \cite{Hatefi:2010ik,Tseytlin:1999dj}. The bosonic effective action for a single D$p$-brane was developed by \cite{Leigh:1989jq}, and extensions to coincident D-branes, including non-Abelian interactions, were explored by \cite{RMyers:1999ps12,Taylor:1999pr}. Supersymmetric formulations and related geometric structures have also been studied in\cite{Howe:2006rv,Cederwall:1996pv}.

While the DBI and CS descriptions capture key features of D-brane dynamics, they omit their higher-derivative corrections that are relevent for effective interactions beyond the leading low-energy approximation. These corrections can be systematically explored by evaluating string scattering amplitudes and expanding them in terms of \(\alpha'\) which is the inverse string tension.

Significant progress has been made in extracting higher-derivative corrections to the effective actions of both stable and unstable D-branes. For BPS D-branes, the four-gauge-field interactions were analyzed to all orders in the \(\alpha'\) expansion in \cite{Hatefi:2010ik}, revealing consistency through order \(\alpha'^4\) with previous results \cite{Barreiro:2005hv}. Further studies obtained higher-derivative interactions involving two gauge fields and two transverse scalars, and developed a universal method for organizing corrections involving four covariant derivatives of transverse scalars in \cite{Hatefi:2012ve}, providing key insights for reconstructing effective interactions of BPS D-branes from their string amplitudes.
Developments on non-BPS D-branes and brane-antibrane systems have been made, particularly regarding the effective interactions of unstable branes as carried out by \cite{Hatefi:2012wj}. Higher-derivative corrections for two gauge fields and two tachyonic fields on brane-antibrane systems were derived, along with other effective actions and their $\alpha'$ corrections in \cite{Hatefi:2012cp}.

A complete review of the Wess-Zumino (WZ) effective action and their generalizations can be found in Section 5 of \cite{Hatefi:2012wj} and \cite{Li:1995pq,Douglas:1995bn, Green:1996dd}. The determination of higher-point open-closed string amplitudes is also crucial for gauge/gravity duality . Understanding the links between open-string excitations and closed-string background fields requires knowledge of D-brane world-volume interactions \cite{Hatefi:2012bp}, as leading-order effective interactions are often inadequate.

Three main methods for deriving effective couplings have been identified in previous analyses of amplitudes with transverse scalar fields and RR vertex operators: non-Abelian Myers interactions, the pull-back of background fields to the D-brane world volume, and the Taylor expansion of bulk fields in transverse fluctuations \cite{RMyers:1999ps12,Hatefi:2012wj}.
New couplings do not appear in Myers terms or Taylor expansions, suggesting that the conventional pull-back prescription needs to be extended, as noted in \cite{Hatefi:2012ve}.

Previous research on higher-order effective interactions, such as \cite{Keurentjes:2004tu} also provides valuable comparisons for our study. We also focus on the dielectric effect and its generalizations. The mixed amplitude with one RR closed-string state, one gauge field, and two transverse scalar fields is key for understanding D-brane polarization in background RR fields. This mechanism is  also important for the dynamics of extended objects in string theory and M-theory, including connections to the \( N^3 \) scaling of M5-brane degrees of freedom explored in \cite{Hatefi:2012ab} and additional M-theory applications in \cite{McOrist:2012yc}. The new couplings derived here may help study higher-derivative corrections to dielectric phenomena and D-brane dynamics \cite{Hatefi:2012bp}.

The dielectric effect is vital for resolving singular geometries in gauge/gravity duality, as RR fluxes can polarize D3-brane configurations into higher-dimensional branes, affecting supergravity backgrounds \cite{Polchinski:2000uf}. Related mechanisms have been examined in M-theory, including D-brane stabilization in curved backgrounds \cite{Bachas:2000ik}, Wess-Zumino-Witten models, fuzzy gravitons \cite{Li:2000bd}, and studies on Kähler manifolds \cite{Douglas:1997sm}, as well as Matrix theory and M-theory compactifications \cite{Hull:1997jc}. Stability conditions have also been investigated in \cite{deAlwis:2013gka}. We recommend \cite{PolchinskiBook} for further details. These findings highlight the need for a deeper understanding of RR-induced interactions beyond leading-order effective action.

The integration techniques discussed in \cite{Fotopoulos:2001pt} identified previously unknown world-volume interactions. Evaluating world-sheet correlation functions with spin fields, fermionic operators, and bilinear currents is also done through the Wick-like prescription from \cite{Liu:2001qa}which is generalized in \cite{Hatefi:2010ik} and provides an efficient method for computing these correlation functions.

In this work, we establish a framework for systematically organizing RR-gauge-scalar interactions by introducing a higher-class pullback operator \( P_{\infty} \) and mixed transgression operators \( \mathcal{T}_{\infty} \). It keeps amplitude-derived pole residues and local contact terms as key physical data, aiming to distinguish admissible tensor structures from coefficients fixed by string-amplitude matching.

We present conjectures related to cohomology and characteristic forms, outlining their validity and consistency conditions. We propose three conjectures regarding the finiteness of the effective operator quotient at fixed derivative weight, the organization of mixed couplings, and the existence of a finite number of primitive tensor structures in the all-order contact terms. We will use explicit S-matrix calculations to assess the proposed geometric organization.

\section{Geometric setting and conventions}

We let $M$ be ten-dimensional spacetime so that
\begin{equation}
 I:\Sigma_{p+1}\hookrightarrow M
\end{equation}
be the embedding of a D$p$-brane world-volume.  We use $a,b,c,\ldots$ for world volume indices and $i,j,k,\ldots$ for transverse directions of D-branes.  Let $E\to\Sigma_{p+1}$ be the Chan-Paton bundle, $A=A_a d\sigma^a$ a connection and

\begin{equation}
 F=dA-iA\wedge A,
 \qquad
 F_{ab}=\partial_aA_b-\partial_bA_a-i[A_a,A_b].
 \label{eq:F}
\end{equation}
The transverse scalar $\Phi^i$ is an adjoint-valued section of $N\Sigma$, with
\begin{equation}
 D_a\Phi^i=\partial_a\Phi^i-i[A_a,\Phi^i].
 \label{eq:Dphi}
\end{equation}

The geometric setting follows the standard differential geometry of vector bundles, connections, and curvature developed in \cite{S. Kobayashi1963}. This formulation provides the natural setting for the world-volume gauge connection and the covariant derivatives of transverse scalar fields.
For an RR potential $C=\sum_q C_q$, the conventional Wess--Zumino action is described as
\begin{equation}
 S_{\mathrm{WZ}}
 =\mu_p\int_{\Sigma_{p+1}}
 \STr\!\left(P[C]\wedge e^{2i\pi F}\right),
 \label{eq:WZ}
\end{equation}

which is embedded in the non-Abelian theory by Taylor and dielectric terms.
\vskip.1in

For a $q$-form $C_q$, the ordinary pull-back is defined as
\begin{equation}
 \big(P[C_q]\big)_{a_1\cdots a_q}
 =C_{\mu_1\cdots\mu_q}(X)
 \partial_{a_1}X^{\mu_1}\cdots\partial_{a_q}X^{\mu_q}.
 \label{eq:pullback}
\end{equation}

Note that in static gauge, transverse fluctuations take shape of \( D_a\Phi^i \), while Taylor expansion introduces normal derivatives \( \partial_i C \). We identify three cases. 1) World-volume form degree   Must equal \( p+1 \) in integrated terms. 2) Normal-class degree which counts \( \Phi \), \( D\Phi \), and normal derivatives of RR fields.
3) Derivative weight  can track powers of Mandelstam invariants and the \(\alpha'\) expansion.

\section{ The  Complete S-Matrix  Analysis}

Our method is based on constructing the complete form of the S-matrix at all orders, gauge covariance, RR gauge transformations, exchange symmetries, and Bianchi identities. The exact amplitude can be decomposed as follows:
\begin{equation}
 \mathcal A_{CA\Phi\Phi}(s,t,u)
 =\mathcal A_{\mathrm{poles}}(s,t,u)
 +\mathcal A_{\mathrm{poles}}(s+t+u)
 +\mathcal A_{\mathrm{contact}}(s,t,u).
 \label{eq:pole-contact}
\end{equation}

The complete amplitude, beyond its pole structure, systematically reveals previously unrecognized RR-gauge-scalar interactions and their higher-derivative extensions.

This work also provides a framework for exploring higher-derivative RR interactions, dielectric polarization effects, and the realization of open-closed string dynamics on the world volume. The standard description of RR interactions involves pullbacks of differential forms, world-volume gauge curvature, transverse Taylor expansions, and non-Abelian dielectric terms for coincident branes. 

 A particular amplitude computation is considered in Ref.~\cite{Hatefi:2012zh} to be known as 
\begin{equation}
 \left\langle V_C V_A V_{\Phi} V_{\Phi}\right\rangle.
 \label{eq:amplitude-symbol}
\end{equation}

This involves one closed-string RR, one gauge field, and two transverse scalar fields. The low-energy expansion involves infinite sequences of massless gauge- and scalar-channel poles, along with local contact interactions across various RR rank sectors. Some interactions are not captured by standard Myers, Taylor, and pullback prescriptions. Thus, we aim to provide a mathematical structure that can accommodate their tensorial and factorization properties.

The transverse components of the RR momentum, denoted \( p^i \) and \( p^j \) appear in the full amplitude and provide couplings that can not be obtained from T-duality transformation. Therefore, a satisfactory classification must retain both types of momenta of the world volume and transverse derivatives of the branes. Fermionic amplitudes  and tachyonic interactions in D-brane–anti-D-brane systems have been studied in Refs. \cite{Antonelli:2019pzg, Hatefi:2019rxq} that provide a complementary methodological approach to this work.

For completeness, we recommend studies of BPS effective actions \cite{Hatefi:2015ora}, open strings \cite{Sagnotti:1987tw,Bianchi:1990yu}, and non-BPS and D-brane anti-D-brane actions \cite{Sagnotti:1992qw,Angelantonj:2002ct}, along with their all-order corrections \cite{Hatefi:2013yxa,Hatefi:2017ags}.
We developed techniques for the five-point amplitude and found closed-form algebraic solutions using Gamma functions, eliminating hypergeometric functions. For details, see Appendix B of \cite{Hatefi:2012wj} for exploring the upper half-plane integrals. The simplified and final version of the amplitude is given by
\begin{equation}
{\cal A}^{C A\phi\phi}={\cal A}_{1}+{\cal A}_{2}+{\cal A}_{3}+{\cal A}_{4},
\end{equation}
where
\begin{align}
{\cal A}_{1}&=&2^{-1/2}4\xi_{1a}\xi_{2i}\xi_{3j}(t+s+u)L_1'
\bigg[k_{3c}k_{2b}\Tr(P_{-}\fsH_{(n)}M_p\Gamma^{jciba})-k_{2b}p^j\Tr(P_{-}\fsH_{(n)}M_p\Gamma^{iba})
\nonumber\\&&-k_{3c}p^i\Tr(P_{-}\fsH_{(n)}M_p\Gamma^{jca})+p^ip^j\Tr(P_{-}\fsH_{(n)}M_p\gamma^{a})\bigg]
\nonumber
\end{align}
The second part of the amplitude is given by
\begin{align}
{\cal A}_{2}&=&2^{-1/2}L_2'
\bigg\{-2us\xi_{1}.k_{2}\xi_{2i}k_{3c}\xi_{3j}\Tr(P_{-}\fsH_{(n)}M_p \Gamma^{jci})-ust\xi_{1a}\xi_{2i}\xi_{3j}\Tr(P_{-}\fsH_{(n)}M_p \Gamma^{jia}) \nonumber\\&&
+2ut k_{3}.\xi_{1}k_{2b}\xi_{3j}\xi_{2i}\Tr(P_{-}\fsH_{(n)}M_p \Gamma^{jib})
+2us \xi_{2i}p^j\xi_1.k_2\xi_{3j}\Tr(P_{-}\fsH_{(n)}M_p\gamma^i)
\nonumber\\&&
-2ut\xi_{3j}p^i\xi_1.k_3\xi_{2i}\Tr(P_{-}\fsH_{(n)}M_p\gamma^j)
\bigg\}\nonumber\end{align}
The third and fourth parts of the amplitude are given by:
\begin{align}
{\cal A}_{3}=2^{1/2}\Tr(P_{-}\fsH_{(n)}M_p\gamma^a)\xi_3.\xi_2 L_1'\bigg[ts\xi_{1a}
-2tk_3.\xi_{1}k_{2a}
-2sk_2.\xi_{1}k_{3a}\bigg]
\nonumber
\end{align}
 
\begin{align}
{\cal A}_{4}=2^{-1/2}L_2'
\bigg\{ 2st\xi_{3}.\xi_{2}k_{2b}k_{3c}\xi_{1a}\Tr(P_{-}\fsH_{(n)}M_p \Gamma^{cba})\bigg\}.
\nonumber
\end{align}

The functions $L_1',L_2'$  defined as
\begin{equation}
L_1'=(2)^{-2(t+s+u)-1}\pi{\frac{\Gamma(-u+\frac{1}{2})
\Gamma(-s+\frac{1}{2})\Gamma(-t+\frac{1}{2})\Gamma(-t-s-u)}
{\Gamma(-u-t+1)\Gamma(-t-s+1)\Gamma(-s-u+1)}}\nonumber
\end{equation}
\begin{equation}
L_2'=(2)^{-2(t+s+u)}\pi{\frac{\Gamma(-u)
\Gamma(-s)\Gamma(-t)\Gamma(-t-s-u+\frac{1}{2})}{\Gamma(-u-t+1)
\Gamma(-t-s+1)\Gamma(-s-u+1)}}
\nonumber
\end{equation}

where  $s+t+u=-p_ap^a.$
For the scalar $t$-channel, we use the field-theory factorization
\begin{equation}
 \mathcal A_t
 =V^i_{\alpha}(C_{p+1},\Phi_3,\Phi)
 G^{ij}_{\alpha\beta}(\Phi)
 V^j_{\beta}(\Phi,A_1,\Phi_2).
 \label{eq:t-factor}
\end{equation}

The scalar kinetic term determines the propagator and the open scalar-gauge vertex, meaning the infinite series in the residue must originate from the RR-scalar-scalar vertex. The effective field theory constructs the all-order scalar coupling as follows:

\begin{align}
 S_{\Phi\Phi C}^{(\infty)}
 =&\;\frac{\mu_p(2\pi\ap)^2}{2(p+1)!}
 \int d^{p+1}\sigma\,
 \epsilon^{a_0\cdots a_p}
 \sum_{n=-1}^{\infty}b_n(\ap)^{n+1}
 \nonumber\\
 &\times\Big[
 \Tr\!\left(
 D_{a_1}\cdots D_{a_{n+1}}\Phi^j
 D^{a_1}\cdots D^{a_{n+1}}\Phi^i
 \right)\partial_jH^{(p+2)}_{i a_0\cdots a_p}
 \nonumber\\
 &\qquad +(p+1)\Tr\!\left(
 \partial_{a_0}D_{a_1}\cdots D_{a_{n+1}}\Phi^j
 D^{a_1}\cdots D^{a_{n+1}}\Phi^i
 \right)H^{(p+2)}_{ij a_1\cdots a_p}
 \Big].
 \label{eq:scalar-allorder}
\end{align}
The index shift ensures that the \( n=-1 \) term reproduces the leading coupling. In the derived three-point vertex, commutator terms from the scalar covariant derivatives do not contribute, which is consistent with the effective field theory and S-matrix analysis.

\subsection{The $p-2=n$ sector}
For $p-2=n$, the amplitude contains infinite $u$-channel gauge poles and new contact interactions.  A new mixed Chern-Simons and pull-back structure is defined as 
\begin{equation}
 S_{C F\Phi\Phi}
 =i(2\pi\alpha')^3\,\mu_p
 \int_{\Sigma_{p+1}}
 \Tr\!\left(
 C_{p-3}\wedge F\wedge D\Phi^i\wedge D\Phi_i
 \right).
 \label{eq:new-wedge}
\end{equation}

Its form degree sums to \(p+1\), making the coupling naturally integrable on the D\(p\)-brane world volume. This structure and its higher-derivative corrections are identified in \cite{Hatefi:2012zh} through direct string amplitude and its matching with effective field theory.

\subsection{The $p=n$ sector}
For $p=n$, the contact terms of the amplitude contain the structure as follows 
\begin{align}
 \mathcal A_{CA\Phi\Phi}^{(p=n)}
 \sim&\;
 \epsilon^{a_0\cdots a_{p-1}a}
 \xi_{1a}\xi_{2i}\xi_{3j}
 \Big(
 p^ip^jH_{a_0\cdots a_{p-1}}
 +p(p-1)k_{3a_0}k_{2a_1}H^{ij}{}_{a_2\cdots a_{p-1}}
 \nonumber\\
 &\hspace{2.8cm}
 -p k_{2a_0}p^jH^i{}_{a_1\cdots a_{p-1}}
 -p k_{3a_0}p^iH^j{}_{a_1\cdots a_{p-1}}
 \Big)\,\mathcal P(s,t,u),
 \label{eq:pn-amp}
\end{align}
where $\mathcal P$ is the all-order contact polynomial.  At leading order, the pull-back and Taylor pieces include
\begin{align}
 S_1=&\;\frac{(2\pi\ap)^3\mu_p}{4(p-1)!}
 \int d^{p+1}\sigma\,\epsilon^{a_0\cdots a_p}
 \Big[
 (p-2)(p-1)\STr\!\left(
 F_{a_0a_1}D_{a_2}\Phi^iD_{a_3}\Phi^j
 \right)C_{ij a_4\cdots a_p}
 \nonumber\\
 &\hspace{2.5cm}
 +2(p-1)\STr\!\left(
 F_{a_0a_1}D_{a_2}\Phi^i\Phi^j
 \right)\partial_jC_{i a_3\cdots a_p}
 \Big],
 \label{eq:S9}
\\[2mm]
 S_{2}=&\;\frac{(2\pi\ap)^3\mu_p}{4(p-1)!}
 \int d^{p+1}\sigma\,\epsilon^{a_0\cdots a_p}
 \STr\!\left(F_{a_0a_1}\Phi^i\Phi^j\right)
 \partial_i\partial_j C_{a_2\cdots a_p}.
 \label{eq:S10}
\end{align}

Compared with direct S-matrix the leading contact term can be written in terms of the RR field strength as
\begin{align}
 S_{3}=&\;\frac{(2\pi\ap)^3\mu_p}{2p!}
 \int d^{p+1}\sigma\,\epsilon^{a_0\cdots a_p}
 \Big[
 p(p-1)\STr\!\left(A_{a_0}\partial_{a_1}\Phi^i\partial_{a_2}\Phi^j\right)
 H_{ij a_3\cdots a_p}
 \nonumber\\
 &\quad +2p\STr\!\left(A_{a_0}\partial_{a_1}\Phi^j\Phi^i\right)
 \partial_iH_{j a_2\cdots a_p}
 +\STr\!\left(A_{a_0}\Phi^i\Phi^j\right)
 \partial_i\partial_jH_{a_1\cdots a_p}
 \Big].
 \label{eq:S11}
\end{align}

where the commutator term in the scalar covariant derivative is  omitted for producing these specific three-open-string couplings.

\subsection{The $p+2=n$ sector}
For $p+2=n$, the amplitude contains non-zero gauge-scalar-scalar-RR contact interactions even though all world-volume indices are saturated.  A leading coupling obtained after combining pull-back and Taylor terms as follows
\begin{equation}
 S_{A\Phi\Phi H}
 =\frac{i(2\pi\ap)^2\mu_p}{p!}
 \int d^{p+1}\sigma\,
 \epsilon^{a_0\cdots a_p}
 \Tr\!\left(A_{a_0}\Phi^j\Phi^i\right)
 H_{ij a_1\cdots a_p}.
 \label{eq:pplus2-leading}
\end{equation}

Higher-derivative corrections and additional new couplings can be derived that are not generated by pull-back, Taylor expansion, or Myers terms. We now construct the higher-class deformation of pull-back, using the full S-matrix analysis and its all-order corrections.

\section{Higher-class deformation of pull-back}

We now construct a higher-class deformation of the pull-back, allowing all functions and their derivatives to be embedded in a closed form. Let us introduce our first proposed mathematical structure for the generalized pull-back as follows.  let \( J^r(N\Sigma \otimes \operatorname{ad}E) \) denote the \( r \)-class bundle of the transverse scalar. We define the formal effective pull-back as follows.
\begin{equation}
 \Pinf[C]
 =P[C]+\sum_{r=1}^{\infty}(\ap)^r P^{(r)}[C].
 \label{eq:Pinf}
\end{equation}
Each $P^{(r)}$ is a local multilinear differential operator built from $C$, its normal derivatives, $F$, $D\Phi$ and higher covariant derivatives.  Locally,
\begin{equation}
 P^{(r)}[C]
 =\sum_I c_{r,I}\,
 \operatorname{Contr}_I\!\left(
 \nabla_{\!\perp}^{q}C,
 D^{m_1}\Phi,\ldots,D^{m_k}\Phi,
 D^{\ell_1}F,\ldots
 \right).
 \label{eq:Pr}
\end{equation}
This is not the pull-back induced by an ordinary map of manifolds.  It is an effective local operator whose coefficients are to be fixed by amplitudes.

An ordinary pull-back obeys $dP[C]=P[dC]$.  We allow the deformed operation to satisfy instead
\begin{equation}
 d\Pinf[C]-\Pinf[dC]
 =\cD_{\infty}(C;F,\Phi),
 \label{eq:transgression-obstruction}
\end{equation}
where \( \cD_{\infty} \) is a gauge-covariant transgression series. RR gauge invariance implies that the integrated term cancels against additional mixed terms or becomes exact. Next, we consider selected Chern-- Simons transgressions as follows.

\section{Selected Chern--Simons transgressions}

We construct mixed transgression operators using the theory of forms and secondary geometric invariants developed by Chern and Simons in \cite{Chern 1974}. The present formulation extends this geometric perspective to mixed RR–gauge–scalar interactions and their higher-derivative corrections. We now let $\hat A$ be the local graded algebra generated by $C$, $H=dC$, $F$, $\Phi$, $D\Phi$ and their covariant derivatives, modulo cyclicity of the trace.   We introduce a selected derivative as
\begin{equation}
 \hat F^0\hat A\subset
 \hat F^1\hat A\subset\cdots,
 \qquad
 \hat A=\bigcup_{r\ge0}\ F^r\hat A.
 \label{eq:filtration}
\end{equation}
We write the irreducibly mixed sector as
\begin{equation}
 \Tinf(C,F,D\Phi)
 =\sum_{r=0}^{\infty}(\ap)^r\cT_r(C,F,D\Phi).
 \label{eq:Tinf}
\end{equation}
The basic $p-2=n$ generator is
\begin{equation}
 \cT_0^{(p-2)}
 =\kappa_0\,
 C_{p-3}\wedge F\wedge D\Phi^i\wedge D\Phi_i.
 \label{eq:T0}
\end{equation}

Higher descendants are obtained by distributing contracted derivatives among $F$ and the scalar one-forms.  A compact and closed form for all-order family is given by
\begin{equation}
 \cT_{\infty}^{(p-2)}
 =\sum_{r,I}\kappa_{r,I}(\ap)^r
 C_{p-3}\wedge
 \mathcal B_{r,I}\!\left(F,D\Phi^i,D\Phi_i\right),
 \label{eq:TB}
\end{equation}

where in the differential operators $\mathcal B_{r,I}$ form a reduced basis considering integration by parts, Bianchi identities, and on-shell trivial terms. At the highest-derivative level, covariant derivatives can be treated as commuting variables, with commutators yielding lower-class corrections. The graded algebra maps momentum polynomials in $s,t,u$ to symmetric symbols, setting up the first stage of amplitude matching.

\subsection{Operator reconstruction from the $\ap$ expansion}

At fixed derivative weight $r$, we choose a basis as follows
\begin{equation}
 \left\{\cO_{r,1},\ldots,\cO_{r,N_r}\right\}
 \subset \hat F^r\hat A
\end{equation}

Having required the field content and total form degree we reduce the basis using integration by parts, Bianchi identities, trace cyclicity, and RR tensor symmetries. Evaluating the remaining operators on plane waves gives rise to a finite linear map  as follows

\begin{equation}
 M_r:\mathbb C^{N_r}\longrightarrow \mathbb C^{K_r},
 \qquad
 M_r x_r=a_r,
 \label{eq:linear-match}
\end{equation}
where $a_r$ is the vector of Mandelstam coefficients extracted from the amplitude.   If $D_{(i)}$ denotes a derivative acting on the $i$th external field, then one can reconstruct Mandelstam variables as 
\begin{align}
 s&\longrightarrow -\frac{\ap}{2}(D_{(1)}+D_{(3)})^2,
 &t&\longrightarrow -\frac{\ap}{2}(D_{(1)}+D_{(2)})^2,
 &u&\longrightarrow -\frac{\ap}{2}(D_{(2)}+D_{(3)})^2.
 \label{eq:symbol-map}
\end{align}

The locations of derivatives are set by solving \eqref{eq:linear-match} in the reduced operator basis. If the kernel of \( M_r \) is non-trivial (\(\ker M_r \neq 0\)) then the amplitude does not uniquely define an off-shell Lagrangian. These kernel directions relate to structures missed by the amplitude, highlighting the need for higher-point amplitudes to resolve ambiguities in derivative ordering and commutators.

\section{A proposed class Chern character}

The new coupling \eqref{eq:new-wedge}  may suggest a compact geometric form that is motivated by the theory of characteristic classes and differential forms developed by Bott and Tu \cite{Bott1}, the index-theory framework of Atiyah and Singer \cite{Atiyah}, and Quillen's theory of superconnections  as well as generalized Chern characters in \cite{Quillen}. We would like to suggest a compact and closed form. Let us introduce a formal effective even curvature as below 
\begin{equation}
 \cF_{\mathrm{eff}}
 =F+\sum_{r=1}^{\infty}(\ap)^r\Xi_2^{(r)}
   +\sum_{r=1}^{\infty}(\ap)^r\Xi_4^{(r)}+\cdots,
 \label{eq:Feff}
\end{equation}

  where $\Xi_{2k}^{(r)}$ are adjoint-valued $2k$-forms built from scalar classs and derivatives of $F$.  At the first nontrivial level one may take
\begin{equation}
 \Xi_2^{(1)}
 =\lambda_1 D\Phi^i\wedge D\Phi_i+\cdots.
 \label{eq:Xi}
\end{equation}
We define the formal class Chern character
\begin{equation}
 \operatorname{ch}_{\mathrm{class}}(F,\Phi)
 =\Tr\exp\!\left(\frac{i}{2\pi}\cF_{\mathrm{eff}}\right).
 \label{eq:chclass}
\end{equation}
Expanding \eqref{eq:chclass} produces the cross term
\begin{equation}
 \Tr\!\left(F\wedge\Xi_2^{(1)}\right)
 \supset
 \lambda_1\Tr\!\left(
 F\wedge D\Phi^i\wedge D\Phi_i
 \right),
 \label{eq:cross}
\end{equation}

 If we multiply the above term by \(C_{p-3}\), then it will have the tensor structure of \eqref{eq:new-wedge}. Equation \eqref{eq:chclass} can be taken as a proposed generating function, not as the fact that \(\cF_{\mathrm{eff}}\) is the curvature of a standard connection. Note that in all the above couplings, non-Abelian ordering must be determined through amplitude matching  with its effective field theory.

\subsection{Normal-class transgression}
The transverse RR momenta appearing in the entire amplitude motivate a separate normal-class operation.  Let $\nabla_{\perp}$ be a connection on $N\Sigma$.  The formal normal Taylor series is
\begin{equation}
 J_{\Phi}C
 =\sum_{q=0}^{\infty}\frac{1}{q!}
 \Phi^{i_1}\cdots\Phi^{i_q}
 \nabla_{i_1}\cdots\nabla_{i_q}C.
 \label{eq:Jphi}
\end{equation}
We define a mixed normal-class transgression by
\begin{equation}
 \Theta_r(C;\Phi,F)
 =\sum_I a_{r,I}\,
 \operatorname{Contr}_I\!\left(
 \nabla_{\perp}^{q}C,
 D^m\Phi,D^n\Phi,D^{\ell}F
 \right),
 \qquad q\ge1.
 \label{eq:theta}
\end{equation}
Unlike a standard Taylor term, \eqref{eq:theta} combines normal derivatives of the RR field with world-volume classes. For plane waves, it generates the products of \(p^i\), \(p^j\), and Mandelstam polynomials that are found in the complete amplitude.

\subsection{Effective coupling cohomology}

 The S-matrix defines local interactions up to integrations by parts and field redefinitions. This construction could be related to the theory of differential characters and geometric invariants developed in \cite{Cheeger}. However, the effective coupling cohomology introduced here is an amplitude-oriented quotient defined through on-shell equivalences, rather than ordinary differential. Let \(\mathcal I_{\mathrm{IBP}}\), \(\mathcal I_{\mathrm{Bianchi}}\), and \(\mathcal I_{\mathrm{EOM}}\) represent the corresponding subspaces, we define

\begin{equation}
 \cH_{\mathrm{eff}}^{r}
 =\frac{\hat F^r\hat A}
 {\hat F^{r-1}\hat A
 +\hat I_{\mathrm{IBP}}
 +\hat I_{\mathrm{Bianchi}}
 +\hat I_{\mathrm{EOM}}}.
 \label{eq:Heff}
\end{equation}

This is an amplitude-oriented quotient that captures independent highest-weight local coupling classes, and this is not de Rham cohomology; rather, a string amplitude defines a class in \eqref{eq:Heff}, with a specific effective Lagrangian as a representative choice. World-volume gauge transformations are
\begin{equation}
 \delta_{\Lambda}A=D\Lambda,
 \qquad
 \delta_{\Lambda}\Phi^i=i[\Lambda,\Phi^i],
 \label{eq:gauge}
\end{equation}

where RR gauge transformations act as \(C \mapsto C + d\Lambda_{\mathrm{RR}}\). Operators from \(F\) and \(D\Phi\) are covariant, while those with \(A\) must form gauge-invariant transgressions and the Ward identity, replacing \(\xi_{1a}\) with \(k_{1a}\) acts as an on-shell test for the gauge field.

\subsection{A compact all-order ansatz}

Combining the previous constructions, we can propose a compact all order ansatz  for the effective action as

\begin{equation}
 S_{\mathrm{eff}}
 =\mu_p\int_{\Sigma_{p+1}}
 \Tr\!\left[
 \Pinf[C]\wedge e^{2i\pi\ap F}
 +\Tinf(C,F,D\Phi)
 \right].
 \label{eq:Seff}
\end{equation}
For the $p-2=n$ sector, it is useful to  show the corrected old vertex and the new mixed sector as follows
\begin{align}
 S_{p-2}
 =&\;\mu_p\int C_{p-3}\wedge
 \left[
 F\wedge F
 +\sum_{r\ge1}(\ap)^r\mathcal K_r(F,F)
 \right]
 \nonumber\\
 &+\mu_p\sum_{r\ge0}(\ap)^r
 \int C_{p-3}\wedge
 \mathcal L_r(F,D\Phi^i,D\Phi_i).
 \label{eq:Spminus2}
\end{align}
The first line is fixed by gauge-pole factorization; the second contains the contact coupling \eqref{eq:new-wedge} and its descendants. For $p=n$ and $p+2=n$, a compact field-strength representation is given by
\begin{equation}
 S_p+S_{p+2}
 =\mu_p\sum_{r\ge0}(\ap)^r
 \int_{\Sigma_{p+1}}
 \Tr\,\cO_r\!\left(
 H,\nabla_{\perp}H,A,F,\Phi,D\Phi
 \right),
 \label{eq:othersectors}
\end{equation}
where the explicit representatives are fixed by the amplitude's component contact terms.

\subsection{Higher operations and derived brackets}
A complementary organization starts from the gauge-covariant exterior derivative $D$ and curvature F, satisfying
\begin{equation}
 D^2X=i[F,X].
 \label{eq:D2}
\end{equation}

We introduce formal multilinear operations \(m_k\) in the algebra, where \(m_2\) represents the ordinary wedge product, and higher operations encode derivative distributions. For example,

\begin{equation}
 m_4(C_{p-3},F,D\Phi^i,D\Phi_i)
 =C_{p-3}\wedge
 \mathcal B(F,D\Phi^i,D\Phi_i)+\cdots,
 \label{eq:m4}
\end{equation}

The omitted terms are lower-class commutator completions. The introduction of higher multilinear operations is motivated by the homotopy-associative structures developed by Stasheff \cite{Stasheff} and the theory of $(A_\infty)$-algebras studied by Getzler and Jones in  \cite{Getzler}. These mathematical frameworks provide possible organizing principles for higher-order products and derivative corrections. We do not claim these operations satisfy full \(A_{\infty}\) or \(L_{\infty}\) identities, but suggest that pole factorization may impose homotopy-like compatibility relations. Higher-point amplitudes like \(CAA\Phi\Phi\) could test this and help resolve commutator ambiguities noted in \cite{Hatefi:2012zh}.

\subsection{Consistency conditions}

We now turn to consistency conditions for the effective couplings. Any proposed all-order coupling must satisfy several constraints as follows:

1) Form Degree: Every integrated term must have degree \(p+1\), for instance, 
\[
\deg C_{p-3} + \deg F + \deg D\Phi^i + \deg D\Phi_i = p + 1.
\]

2) World-Volume Gauge Symmetry: \(A\) should appear through \(F\) and scalars through \(D\Phi\). Explicit \(A\)-dependent terms must have gauge variations that are exact or canceled by companion terms.

3) RR Gauge Symmetry: Under \(C \to C + d\Lambda_{\mathrm{RR}}\), the total variation of \eqref{eq:Seff} must reduce to a boundary term, constraining the mixed transgression series.

4) Exchange Symmetry: Scalar insertions can be split into symmetric and antisymmetric sectors; the operator basis should be decomposed accordingly.

5) Factorization: The residue at every massless pole must equal the product of lower-point vertices and the relevant propagator, determining which \(\ap\) corrections belong to existing vertices.

6) On-Shell Equivalence: Operators differing by denominator elements in \eqref{eq:Heff} represent the same amplitude class; coefficients should be compared after reducing to a common basis. We now turn to various conjectures.

\section{Conjectures and mathematical remarks}

The complete amplitude raises several mathematical questions that higher point functions may be able to address. Let us point them out as follows.

 First Conjecture: For fixed field content and derivative weight, the quotient \(\cH_{\mathrm{eff}}^r\) is finite-dimensional, allowing every order in \(\ap\) to be reconstructed using finite-dimensional linear algebra, despite the full action having infinitely many terms.

Second Conjecture: A nontrivial subset of the all-order mixed RR–gauge–scalar sector is generated, modulo on-shell equivalence, by a finite collection of effective even forms \(\Xi_{2k}^{(r)}\) in a generalized characteristic form like \eqref{eq:chclass}.

Third Conjecture: In the \(CA\Phi\Phi\) contact interaction terms, infinitely many \(\ap\) coefficients multiply only finitely many primitive tensor types, with higher orders generated by derivative-symbol polynomials and lower-class non-Abelian completions.

 To test the third conjecture, we need to construct the reduced operator basis across various derivative weights. If the number of primitive contractions stabilizes while polynomial degree increases, it then supports the third Conjecture. For higher class pull-backs one may make a remark. For embeddings \(\Sigma \to M \to N\), ordinary pull-back is strictly functional, while the effective operation \(P_{\infty}\) may only be functional up to selected transgression, possibly connecting to superconnections or differential K-theory, although no such identification is worked out here. A possible differential-cohomological interpretation is motivated by Freed's formulation of generalized differential cohomology and its role in charge quantization \cite{Freed}.

\subsection{Additional Coupling Families and Explicit Derivative Structures}
The all-orders contact expansions in the amplitude are expressed via coefficient families \(c_n\), \(c_{n,m}\), \(e_{p,n,m}\), and \(f_{p,n,m}\). These can be represented covariantly by differential operators, with coefficients reproducing corresponding Mandelstam polynomial. For instance, a factor \((st)^m\) acting on a field content\(HA\Phi\Phi\) can be represented by

\begin{equation}
 (st)^m HA\Phi\Phi
 \;\longrightarrow\;
 (\ap)^{2m}H\,
 \partial_{a_1}\cdots\partial_{a_{2m}}A\,
 D^{a_1}\cdots D^{a_m}\Phi^i\,
 D^{a_{m+1}}\cdots D^{a_{2m}}\Phi_i,
 \label{eq:stmap}
\end{equation}
up to integrations by parts where the precise distribution is dictated by the tensor structure.  Similarly,
\begin{equation}
 (s+t+u)^{r}HA\Phi\Phi
 \;\longrightarrow\;
 \left(\frac{\ap}{2}\right)^r
 H\,(D_aD^a)^r(A\Phi\Phi).
 \label{eq:sumstu}
\end{equation}

These formulas represent the momentum-to-derivative map, with the exact covariant representative obtained after reduction in \(\mathcal H_{\mathrm{eff}}^r\). Another interaction term for the \(p=n\) contact interaction is given by

\begin{align}
 \mathcal I^{(r)}_{p}
 =&\;\frac{\mu_p(2\pi\ap)^3}{2p!}(\ap)^r
 \int d^{p+1}\sigma\,\epsilon^{a_0\cdots a_p}
 \Big\{
 p(p-1)\,\Tr\!\left[
 \mathfrak D^{(r)}A_{a_0}\,
 \partial_{a_1}\Phi^i\,
 \partial_{a_2}\Phi^j
 \right]H_{ij a_3\cdots a_p}
 \nonumber\\
 &\qquad+2p\,\Tr\!\left[
 \mathfrak{ D}^{(r)}A_{a_0}\,
 \partial_{a_1}\Phi^j\,\Phi^i
 \right]\partial_iH_{j a_2\cdots a_p}
 +\Tr\!\left[
 \mathfrak{ D}^{(r)}A_{a_0}\,\Phi^i\Phi^j
 \right]\partial_i\partial_jH_{a_1\cdots a_p}
 \Big\},
 \label{eq:Ipfamily}
\end{align}

The derivative operators have symbols determined by the polynomial families in the contact terms of the amplitude, clarifying that the same leading tensor structures may correspond to different higher-derivative polynomials. In the \(p+2=n\) case, the amplitude includes a leading term with \(A_{a_0}\) (from Eq.~\eqref{eq:pplus2-leading}) and derivative terms involving \(\partial_bA_a\), \(D_bD_a\Phi^i\), and normal derivatives of \(H\). These can be organized as follows:

\begin{equation}
 S_{p+2}^{\mathrm{contact}}
 =\frac{\mu_p(2\pi\ap)^2}{p!}
 \sum_{r\ge0}(\ap)^r
 \int d^{p+1}\sigma\,
 \epsilon^{a_0\cdots a_p}
 H_{ij a_1\cdots a_p}\,
 \mathcal J_{a_0}^{ij,(r)},
 \label{eq:Jcurrent}
\end{equation}
with a matrix-valued current
\begin{align}
 \mathcal J_{a}^{ij,(0)}&=i\Tr(A_a\Phi^j\Phi^i)\nonumber
\end{align}
\begin{align}
 \mathcal J_{a}^{ij,(1)}=
 \rho_1\Tr(\partial_bA_a D^bD_c\Phi^iD^c\Phi^j)
 +\rho_2\Tr(\partial_bA_aD_c\Phi^iD^cD^b\Phi^j)\\
 +\rho_3\Tr(A_aD_bD_c\Phi^iD^bD^c\Phi^j)+\cdots
\nonumber
\end{align}

The coefficients \(\rho_k\) are determined by low-energy amplitude coefficients in a fixed basis, not as free physical parameters. Thus, Equation~\eqref{eq:Jcurrent} is a compact definition, with component data related to direct matching to the S-matrix elements with effective field theory.

\subsection{Gauge identities and momentum constraints}

The \(p+2=n\) case in the amplitude illustrates how gauge invariance and RR Bianchi identities constrain seemingly independent terms. Hence, the relevant identity has the form:

\begin{equation}
 \epsilon^{a_0\cdots a_p}
 \left[
 -p_{a_p}(p+1)H_{ij a_0\cdots a_{p-1}}
 -p_jH_{i a_0\cdots a_p}
 +p_iH_{j a_0\cdots a_p}
 \right]=0,
 \label{eq:bianchiRR}
\end{equation}

This relation arises from \(dH_{p+2}=0\) and is crucial for verifying the gauge Ward identity after substituting \(\xi_{1a} \to k_{1a}\). In higher-class language, Eq.~\eqref{eq:bianchiRR} shows that normal-class operators are not independent of transverse derivative operators. Therefore, the Bianchi ideal \(\mathcal I_{\mathrm{Bianchi}}\)  must include mixed transverse and world-volume relations, reducing the apparent dimension of the coupling space and preventing overcounting. Note that in three-open-string couplings, the commutator contribution \(D_a\Phi^i = \partial_a\Phi^i - i[A_a, \Phi^i]\) does not contribute to the required vertex and must be replaced by the partial derivative. This is not a universal rule, as higher-point amplitudes can reveal commutator terms hidden in the \(CA\Phi\Phi\) amplitude. 

\subsection{Comparison with ordinary Pull-back, Taylor and Myers terms}

Ordinary geometric prescriptions can be embedded within this notation. The Taylor expansion is the zero-world-volume-derivative part of the normal-class series and one considers

\begin{equation}
 C(X^a,\Phi^i)
 =C(X^a,0)+\Phi^i\partial_iC
 +\frac{1}{2}\Phi^i\Phi^j\partial_i\partial_jC+\cdots.
 \label{eq:Taylor}
\end{equation}

Pull-back provides essential tangent one-forms \(D\Phi^i\), while Myers terms contribute vital interior contractions and scalar commutators. Together, these operators generate crucial leading structures seen in Eqs.~\eqref{eq:S9}--\eqref{eq:S10} and in the \(p+2=n\) case  the mixed pull-back/Taylor combinations are highlighted in Eq.~\eqref{eq:pplus2-leading}.

Once all ordinary pull-back, Taylor, Myers, and higher-derivative corrections are accounted for, any residual regular amplitude must be represented by additional local operators. This concept does not imply that a term cannot be transformed, after integration by parts, into a form that resembles existing structures; rather, it denotes that its coefficient and complete tensor combination cannot be generated only by the three standard Myers, pull-back, and Taylor expansion methods alone.

The higher-class pull-back is intentionally designed to broaden the pull-back sector while preserving this crucial distinction. Thus, we maintain the decomposition, ensuring clarity in our approach as follows 

\begin{equation}
 S_{\mathrm{eff}}=S_{\mathrm{deformed\;pullback}}+S_{\mathrm{mixed\;transgression}}
 \label{eq:decomp}
\end{equation}

which can be taken as part of the definition of the formalism.  Different choices of representatives may shift terms between the two pieces, but their class in the on-shell quotient remains fixed.

\section{Discussion and conclusion}

The complete RR–gauge–scalar–scalar amplitude rigorously tests how conventional geometric prescriptions could shape the D-brane effective action. Its low-energy expansion restricts massless factorization channels and local interactions. In this context, transverse RR momenta hold information about the normal dependence of background fields, while the Mandelstam invariants \(s, t, u\) organize the world-volume derivative expansion and this crucial fact  is essential for the proposed classification.

The framework developed here relates complementary roles to the higher-class pullback \(P_{\infty}\) and the mixed transgression series \(\mathcal T_{\infty}\). \(P_{\infty}\) organizes deformations related to RR insertions and their normal dependence, while \(\mathcal T_{\infty}\) captures mixed RR–gauge–scalar interactions that are not fully addressed by conventional pullback. For example, the differential-form expression is taken as

\begin{equation}
 S_{C F\Phi\Phi}
 =i(2\pi\alpha')^3\,\mu_p
 \int_{\Sigma_{p+1}}
 \Tr\!\left(
 C_{p-3}\wedge F\wedge D\Phi^i\wedge D\Phi_i
 \right).
\nonumber
\end{equation}

This coupling represents an amplitude-constrained family of higher-derivative descendants. The reconstruction procedure is outlined in Eq.~\eqref{eq:linear-match}. At each order in \(\alpha'\), one determines a covariant basis of local operators modulo prescribed on-shell equivalences, evaluates their contributions to relevant amplitudes, and fixes their coefficients by comparing them to the low-energy expansion of the string amplitude. Any compact resummation, including the proposed class Chern character, should be viewed in relation to this matching procedure rather than as an independent determination of effective couplings.

The three conjectures address the finiteness of the effective operator quotient at fixed derivative weight, the potential organization of mixed couplings in characteristic forms, and the existence of finitely many primitive tensor structures in the all-order contact terms    . Establishing these statements and testing their compatibility with higher-point amplitudes and non-Abelian gauge identities are crucial mathematical challenges. Ultimately, explicit S-matrix calculations remain the physical criterion for which this proposed geometric prescription must be assessed.

 Recent advances in using neural networks for quantum perturbation theory and related fields, such as cosmology and numerical relativity, are examined in \cite{Hatefi:2024asc}, \cite{Hatefi:2025lqe}, \cite{Hatefi:2023sgr}, and \cite{Hatefi:2020gis}. We are developing these numerical methods, advanced machine learning and artificial neural network methods to actually study the physics and mathematics of scattering amplitudes \cite{Hatefi:2026} as well.

\section*{Acknowledgments}

Ehsan Hatefi thanks Giulio Bresciani, Dario Trevisan, Giovanni Gaiffi, Enrico Trincherini, Augusto Sagnotti, Luis Alvarez-Gaume, Per Sundell, Kumar Narain, and Nima Arkani-Hamed for their valuable discussions. This work was supported by E.H.'s International Maria Zambrano Grant and INFN. Armin Hatefi acknowledges support from the Natural Sciences and Engineering Research Council of Canada (NSERC). Ehsan Hatefi also appreciates the hospitality of CERN, ICTP, the University of Pisa, Scuola Normale Superiore, Alcalá University, the Perimeter Institute, the Institute for Advanced Study in Princeton, KITP in Santa Barbara, and the Simons Center for Geometry and Physics, where parts of this work took shape during his visits.


\end{document}